\documentclass[a4paper,11pt]{article}
\usepackage{jcappub}

\def\l{\left}
\def\r{\right}

\title{Spherical Collapse in Modified Gravity Theories}

\author[a,b]{Mario A. G\'omez,}
\author[a]{Jorge L. Cervantes-Cota \note{Corresponding author.}}

\affiliation[a]{Instituto Nacional de Investigaciones Nucleares, ININ.\\M\'exico}
\affiliation[b]{Universidad Nacional Aut\'onoma de M\'exico, UNAM.\\M\'exico City, M\'exico}

\emailAdd{mario.gomez@ciencias.unam.mx}
\emailAdd{jorge.cervantes@inin.gob.mx}

\abstract{We study the spherical collapse in the Parametrized Post-Friedmannian (PPF) scheme. We use a general form of the PPF parameter related to the Poisson equation and found the equations to solve that includes a non-trivial fifth force coming from the convolution of the modified gravity term in the \textit{k}-space.  In order to compute a concrete model, we use the parametrization proposed by Bertschinger and Zukin. The equations of the spherical collapse are solved assuming a Gaussian density profile and we show there is no shell crossing before reaching the turn around point. We show that the fifth force does not satisfy the Birkhoff's theorem and introduces different behaviors for the density threshold $\delta_{c}$, which in this case depends on the size and shape of the initial density profile, and therefore one expects a different statistic of the collapsed objects in the universe.}

\begin{document}
\maketitle
\flushbottom

\section{Introduction}

One of the most exciting open  problems in cosmology is the well understanding of the mechanism that carries the observed late-time accelerating expansion of the Universe 
\cite{Riess:1998cb, Perlmutter:1998np}.  Since the first evidence of the need of a new component in the universe called Dark Energy (DE), the introduction of the well known 
cosmological constant ($\Lambda$) as DE has succeed a variety of observations \cite{Ade:2013zuv} and has established the standard cosmological model $\Lambda$CDM, which assumes
General Relativity (GR) and also introduces a dark matter component \cite{Zwicky:1942zz}. However, the cosmological constant leaves doubts about its nature. Hence, many other 
theoretical alternatives have been also proposed that try to explain the accelerating expansion in different ways. One approach is to add an ``exotic matter" source with negative 
presure in the right hand side of the Einstein equation; within this approach one has quintessence \cite{Caldwell:1997ii}, k-essence \cite{Chiba:1999ka}, and perfect fluid models 
\cite{Kamenshchik:2001cp}. The other approach is the called ``modified gravity" (MG) models, that modify the left hand side of the Einstein equations. This includes $f\l(R\r)$ gravity
 \cite{Capozziello:2002rd,Capozziello:2003gx,Carroll:2003wy}, scalar-tensor theories \cite{Amendola:1999qq,Uzan:1999ch,Chiba:1999wt} and braneworld models \cite{Dvali:2000hr,Sahni:2002dx}.\\

\noindent If any of the proposed theories attempt to be valid and to shed light for a better understanding of the dark sector then they must be confronted with observations. Typically, one has to work out the perturbation equations for each theory and incorporate them into an Einstein-Boltzmann solver or a N-body code to construct a list of observables to 
compare to the data. This task can be cumbersome and time-consuming. Over the last years it has been of great interest the developing of an intermediate step between theory and data that allows us to avoid working out each model and encompass a wide range of theoretical schemes, at least at background and first order perturbation level. This phenomenological approach has been dubbed as ``Parametrized Post-Friedmannian" scheme (PPF) \cite{Hu:2007pj}, inspired by the so called Parametrized Post-Newtonian (PPN) formalism \cite{Will:1981cz} created to confront alternative theories of gravity with Solar System measurements. The idea is to assume a $\Lambda$CDM model at background level and to parametrize deviations of GR at first order in perturbations.\\

\noindent There are several modified gravity parametrizations, of which one widely used is the one proposed by Bertschinger and Zukin \cite{Bertschinger:2008zb} (BZ), where two functions of time
and scale are introduced motivated by $f\l(R\r)$ solutions in the quasi-static limit which has been shown in recent works to be valid for large scale structure applications in $f\l(R\r)$ models \cite{Bose:2014zba}. This is the approach we closely used in this work. In addition to this parametrization some other general frameworks have been developed, such as the Hu-Sawicki model \cite{Hu:2007pj}, its relativistic corrections \cite{Lombriser:2013aj} and the quite general approach
proposed by Baker and Ferreira in \cite{Baker:2012zs}. These parametrizations are not independent, but related to each other as mapped en \cite{Daniel:2010ky}.  The target that all 
these different approaches have in common is to create a link between modified gravity theories with its corresponding PPF parameters.\\

\noindent However, the objects from which cosmological information is extracted can be beyond the linear perturbation regime (galaxies and clusters of galaxies). Therefore, the study of gravitational collapse and dynamics of collisionless systems at non-linear perturbations is  necessary to complete the description of structure formation in the Universe. Although, the full non-linear dynamics has to be dealt only with numerical simulations, one can make some assumptions about the system symmetry in order to construct some analytical models that provide a valuable first approximation to the non-linear regime.  One of these analytical models is the so called Spherical Collapse (SC) model introduced in \cite{Gunn:1972sv}.  In this model a spherical, symmetric overdensity evolves to form a virialized bound system via gravitational instability. This model can be applied in different cosmological contexts to study the abundances of virialized objects through the Press-Schechter formalism \cite{Press:1973iz,Peacock:1999} or relaxing the spherical symmetry through its extension of Sheth and Tormen \cite{Sheth:1999mn}.\\

\noindent Lately, many authors have used this model to study the halo mass function and the linear halo bias \cite{Sheth:1999mn} in different cosmological scenarios. Adding the presence of massive neutrinos \cite{LoVerde:2014rxa}, in braneworld cosmology \cite{Schmidt:2009yj}, in Galileon gravity \cite{Barreira:2013xea}, in context of $f\l(R\r)$ theories \cite{Schmidt:2008tn, Kopp:2013lea,Guo:2013dha} or more recently in chameleon theories \cite{Lombriser:2013eza}.  In this work we construct and solve the SC model in the BZ scheme developed in \cite{Bertschinger:2008zb}, but after constructing the formalism one can move on and solve the SC in any other kind of parametrization, which is the main reason of using a parametrized scheme of modified gravity.\\

\noindent This paper is organized as follows. In Section II we present the PPF scheme and the $f\l(R\r)$ case is discussed in Section III. In Section IV we implement a PPF scheme to the
SC Model and we show our main results. Finally, we summarize our findings in Section V.

\section{Parametrized Modified Gravity}

Considering linear scalar perturbation and the perturbed metric tensor in the conformal Newtonian gauge:

\begin{equation}\label{metric}
\mathrm{d}s^{2} = a\left( \eta \right)^{2}\left[-\left(1+2\Psi \right)\mathrm{d}\eta^{2}+\left(1+2\Phi \right)\mathrm{d}x^{2}\right],
\end{equation}

\noindent where $\eta$ refers to the conformal time. The growth of structure in the universe at linear perturbation level is determined by the Poisson and the gravitational slip equations. These two equations in Fourier space read (neglecting any contribution from the anisotropy stress, $\sigma$):
 
\begin{equation}
k^2\Psi = - 4 \pi G a^2 \bar{\rho} \Delta \label{poisson} 
\end{equation}
\begin{equation}
\Phi +\Psi  =0 . \label{slip}
\end{equation}

\noindent where $k$ is the wavenumber, $\Delta \equiv \delta +3 a H (1+\omega)\theta / k^2$ is the comoving density contrast and $a$ is the scale factor. Note that equation (\ref{poisson}) is not the original Poisson equation for the potential $\Phi$ that comes from combining the time-time and the time-space components of Einstein equations. Instead, we have derived an analogous Poisson equation for the potential $\Psi$, which is more closely related to observation. In fact, $\Psi$ is the potential that governs the non-relativistic particles motion.\\

Alternative theories of gravity generate a different cluster dynamics and CMB evolution from the $\Lambda$CDM model. Equations (\ref{poisson}) and (\ref{slip}) have in general different forms, depending on the class of Modified Gravity (MG) theory involved.   Thus, the task to study the evolution of structure and CMB anisotropies in different MG models could be tedious and  cumbersome.  For this reason the PPF formalism was put forward that let us study different theoretical MG models within the same first order perturbation formalism. This can be achieved  by introducing two free functions of time (or scale factor) and scale: $\mu \l(a,k\r)$ and $\gamma \l(a,k\r)$, that modify the Poisson and gravitational slip equations, in order to encompass a group of MG models.  Accordingly, one can write:
  
\begin{equation}
k^2\Psi = - 4 \pi G a^2 \mu \l(a,k\r) \bar{\rho} \Delta \label{poissonmg} 
\end{equation}
\begin{equation}
\Phi + \gamma \l(a,k\r) \Psi  =0 . \label{slipmg}
\end{equation}

\noindent where the usual GR case is recovered when $\mu = \gamma =1$. A general modification of gravity through an arbitrariness in the explicit form of the functions $\mu \l(a,k\r)$ and $\gamma \l(a,k\r)$ could be tedious and difficult to constrain.  Silvestri et. al. \cite{Silvestri:2013ne} have shown that in a set of viable models in which local theories of gravity are considered possessing only one extra degree of freedom and second order equations of motion in the Quasi-Static Limit (QSL), the functions $\mu \l(a,k\r)$ and $\gamma \l(a,k\r)$ must be ratios of even polynomials of second order in $k$ and in general the numerator of $\mu \l(a,k\r)$ is the same as the denominator of $\gamma \l(a,k\r)$.  Introducing five functions of the background $p_{i}\l(a\r)$ one has:

\begin{equation}
\gamma = \frac{p_{1}\left(a\right)+p_{2}\left(a\right)k^{2}}{1+p_{3}\left(a\right)k^{2}}, \label{nppf1}
\end{equation}

\begin{equation}
\mu=\frac{1+p_{3}\left(a\right)k^{2}}{p_{4}\left(a\right)+p_{5}\left(a\right)k^{2}}. \label{nppf2}
\end{equation}

\noindent Although this form of parametrized MG is found in the QSL, in which $k/ a H \gg 1$ is assumed, these expressions are quite suitable for near-horizon scales since near and super-horizon effects have practically no impact on observables in viable MG models when relativistic effects are neglected \cite{Hojjati:2012rf};  for a treatment of relativistic effects see \cite{Lombriser:2013aj}. 

\section{Parametrized MG in $f\left( R \right) $ Theories} 

One of the simplest MG models is the $f\l(R\r)$ gravity in which the 4-dimensional action is given by some general function $f\l( R \r) $ of the Ricci scalar $R$, in the Jordan frame:

\begin{equation}
S = \frac{1}{2 \kappa^2} \int \mathrm{d}^{4} x \sqrt{-g} f \l( R \r) + S_{m} \l( g_{\mu \nu}, \psi_{m} \r), \label{action}
\end{equation}

\noindent where $\kappa^{2}=8\pi G$, and $S_{m}$ is a matter action with minimal coupled matter fields $\psi_{m}$. By varying the action (\ref{action}) with respect the metric one obtains the field equation:

\begin{equation}
f_{R} R_{\mu\nu}-\frac{1}{2} f\l( R \r) g_{\mu\nu}-\nabla_{\mu} \nabla_{\nu}f_{R} + g_{\mu\nu}  \Box f_{R} = \kappa^{2} T_{\mu\nu},
\label{fr1}
\end{equation}

\noindent where $f_{R} \equiv \partial f / \partial R$ and $T_{\mu\nu}$ is the matter energy-momentum tensor. The trace of the above equation is given by
\begin{equation}
 3 \Box f_{R} + f_{R} R - 2 f\l( R \r) =  \kappa^{2} T,
\label{fr2}
\end{equation}

\noindent where $T = g^{\mu\nu}T_{\mu\nu}$. Equation (\ref{fr2}) governs the dynamics of a scalar field $\varphi \equiv f_{R}$, dubbed scalaron \cite{Starobinsky:1980te}, which is a propagating scalar degree 
of freedom. Any viable $f\l(R\r)$ model must satisfy a number of conditions, i.e.,  to avoid instabilities and tachyonic scalaron ($f_{R R} >0$); to evade the existence of ghosts \cite{Nariai:1973eg}
($f_{R}>0$); to recover GR at early times ($f\l(R\r)\rightarrow R-2\Lambda$ for $R\rightarrow \infty$) and at solar system ($| f_{R0}|-1< 10^{-6}$, $f_{R}$ evaluated today \cite{Hu:2007pj}. See \cite{Koyama:2015vza} for a complete discussion  about $f\l(R\r)$ theories.\\

Given an expansion history $ H \l( a \r)$ and $\dot{H} \l( a \r)$, it is possible to design a $f\l(R\r)$ that solves (\ref{fr1}) constrained by the boundary condition $f_{R0}$. Thus,  $f\l(R\r)$ models constructed in this way can have an expansion history practically indistinguishable from the $\Lambda$CDM model \cite{Hu:2007pj,Amendola:2006kh}. To be able to distinguish among MG models, it is then necessary to study the observables of linear and non-linear theory, e.g. the evolution of cosmic structures or the growth factor, preferably at different redshifts, and to compare them with observations.\\

To describe structure formation in a $f\l(R\r)$ model it is necessary to expand $f_{R}$ in perturbation theory, $\delta f_{R}$. Hence, in $f\l(R\r)$ new terms appear in the standard Poisson and gravitational slip equations. One obtains, at first order \cite{Hojjati:2012rf}, 

\begin{eqnarray}
k^{2} \Psi - k^{2} \frac{\delta f_{R}}{2 f_{R}}+\frac{3}{2}\l[ \l( \dot{\mathcal{H}}-\mathcal{H}^{2} \r) \frac{\delta f_{R}}{f_{R}} -\l(\dot{\Phi}-\mathcal{H} \Psi \r) \frac{\dot{f_{R}}}{f_{R}}\r]&=&\frac{a^{2}}{2\kappa^2}  \frac{\rho}{f_{R}} \Delta, \label{PoissonfR}\\
\Psi + \Phi &=& -\frac{\delta f_{R}}{f_{R}} \label{slipfR}.
\end{eqnarray}

As discussed in \cite{Pogosian:2007sw}, in a $f\l(R\r)$ model the Compton wavelength of the scalaron $\lambda_{C} \equiv 2\pi / m_{f_{R}}$, where
\begin{equation}
m_{f_{R}}^{2} \equiv \frac{1}{3}\l(\frac{f_{R}}{f_{RR}}-R\r),
\end{equation}

\noindent is the effective scalaron mass, that sets two regimes in the dynamics. In scales above $\lambda_{c}$ modifications are negligible and GR is recovered. On the other hand, in scales smaller than the Compton wavelength the growth is enhanced by a fifth force and in addition the metric potentials are no longer equal. Also, if one assumes the QSL, the two above equations can be reproduced by equations (\ref{poissonmg}) and (\ref{slipmg}) with a suitable $\mu$ and $\gamma$:

\begin{eqnarray}
\mu^{Q} &=& \frac{1}{f_{R}}\frac{1+\l(4/3\r)Q}{1+Q}  ,\label{muq}\\
\gamma^{Q} &=& \frac{1+\l(2/3\r)Q}{1+\l(4/3\r)Q} ,\label{gammaq}
\end{eqnarray}

\noindent in which $Q = 3 \frac{k^{2}}{a^{2}} \frac{f_{RR}}{f_{R}}$.  Inspire by this in \cite{Bertschinger:2008zb} it is proposed the following parametrization for $\mu$ and $\gamma$:

\begin{eqnarray}
\mu^{BZ} &=&\frac{1+\frac{4}{3}\lambda_{1}^{2} k^2 a^{s}}{1+\lambda_{1}^{2} k^2 a^{s}}  ,\label{mubz}\\
\gamma^{BZ} &=&\frac{1+\frac{2}{3}\lambda_{1}^{2} k^2 a^{s}}{1+\frac{4}{3}\lambda_{1}^{2} k^2 a^{s}}  ,\label{gammabz}
\end{eqnarray}

\noindent where $\lambda_{1}$ represents the Compton wavelength and $s$ encodes its time dependence.  In \cite{Hojjati:2012rf} it is shown that this last form of parametrizing with $s=4$ is quite accurate and can be safety used for deriving constraints on $f\l(R\r)$ models for next generation of large scale surveys.  Notice that (\ref{mubz}) and (\ref{gammabz}) fall within the more general form of  $\mu$ and $\gamma$ discussed before, eqs. (\ref{nppf1}) and (\ref{nppf2}).

\section{Spherical Collapse within the PPF formalism}

In this section we will develop a semi-analytic SC model considering a MG force due the modified Poisson equation (\ref{poissonmg}) where $\mu\l(a,k\r)$ will be given by equation (\ref{nppf2}). The SC model treats the evolution of thin, adjacent concentric mass shells forming a constant spherical symmetric perturbation in the cosmic density field. See \cite{white:2008} for a standard treatment of SC model.\\

\noindent In the Newtonian limit the evolution of each shell is governed by the Newtonian force law. Considering dark matter and dark energy components one has:

\begin{equation}
\frac{\mathrm{d}^{2}r}{\mathrm{d}t^{2}} = - \nabla \Psi - \frac{1}{2} H^{2} \Omega_m r + H^{2} \Omega_{\Lambda} r, \label{newton}
\end{equation}

\noindent where $r$ is the physical position of inner shells. $\Omega_{m}$ and $\Omega_{\Lambda}$ are the dimensionless matter and cosmological constant energy densities, respectively.  Usually, the SC model considers a constant, or top-hat, initial density profile, which is saved from shell crossing in both Einstein de-Sitter (EdS) and $\Lambda$CDM model. Similar as it happens in $f\l( R \r)$ models, an initial top-hat profile suffers from shell-crossing generating a large spike near the perturbation edge \cite{Kopp:2013lea}.  To avoid shell crossing, in general, a perturbation with a monotonically decreasing $\bar{\delta_{i}}$ profile is needed:

\begin{equation}
\bar{\delta}\l(r\r) = \frac{3}{r^{3}}\int_{0}^{r} \delta \l(r'\r)r'^{2} \mathrm{d} r'.
\end{equation}

\noindent where the bar denotes volumen average. For this reason we will consider an initial Gaussian profile for the density profile. This alleviates the shell crossing and incorporates a more realistic perturbation,

\begin{equation}
\rho_{i}\l(r\r) = \rho_{0} \exp\left(-\lambda^{2}r^{2}\right),\label{gauss1}
\end{equation}

\noindent where  ${\rho_{0}}$ is the initial magnitude of the density profile at $r=0$. Before proceeding to solve (\ref{newton}) one has to calculate the explicit form of the force term $-\nabla \Psi$, which is the only place a MG model is captured, assuming a standard $\Lambda$CDM background dynamics as in the PPF formalism; otherwise $H$ also encodes a different background dynamics.\\

\noindent Firstly, we have to transform the Poisson equation (\ref{poissonmg}) into the configuration space, in the QSL we have:

\begin{eqnarray}\label{rn1}
\nabla^{2} \Psi =  4 \pi G a^{2} \l(\delta \rho \ast  \mathcal{F}^{-1}  \mu  \r), \label{possionreal}
\end{eqnarray}

\noindent $\mathcal{F}^{-1}$ means the inverse Fourier transform and ``$\ast$" denotes the convolution between the density contrast and the $\mu$ function. Assuming (\ref{nppf2}), we obtain:

\begin{eqnarray}\nonumber
 \mathcal{F}^{-1}  \mu \equiv \mu \l( r \r) &=& \frac{1}{4 \pi} \frac{p_{4}}{p_{5}}\l(\frac{1}{p_{4}}-\frac{p_{3}}{p_{5}}\r) \frac{e ^{-\sqrt{\frac{p_{4}}{p_{5}}}r }}{r} +A \delta_{D}\l(r\r)\\ \label{mu}
&=& \frac{1}{4 \pi} \beta^{2} \alpha  \frac{e ^{-\beta r }}{r} +A \delta_{D}\l(r\r),
\end{eqnarray}

\noindent where $\delta_{D}$ denotes de Dirac delta and

\begin{eqnarray}
\alpha & \equiv & \frac{1}{p_{4}}-\frac{p_3}{p_5},\\
\beta^2  & \equiv & \frac{p_{4}}{p_{5}},\\
A &\equiv& \lim _{k \rightarrow \infty} \mu = \frac{p_{3}\l(a\r)}{p_{5}\l(a\r)}.
\end{eqnarray}

\noindent Expanding the convolution in the Poisson equation, we finally get

\begin{eqnarray}
\nabla^{2} \Psi =  4 \pi G \left[ \frac{1}{2} \int_{0}^{R_{i}} \chi \l(r,r'\r)\l(\frac{\mathrm{d}M}{\mathrm{d}r'}\r)\mathrm{d}r' + A \rho - \l(\alpha +A \r) \bar{\rho} \right]  . \label{Poissonexp }
\end{eqnarray}

\noindent where we have defined $\chi \l(r,r'\r) \equiv \int_{0}^{\pi}F\l(|\mathbf{r}-\mathbf{r'}|\r) \sin \l(\theta\r)\mathrm{d}\theta$, $F\l(r\r)\equiv \frac{1}{4 \pi} \beta^{2} \alpha  \frac{e ^{-\beta r }}{r}$ and $M\l(r\r)$ is the mass function that provides the total mass inside a radius $r$. Integrating this last expression to obtain the wanted force expression:

\begin{equation}
-\nabla \Psi = - \frac{4 \pi G}{r^2} \int_{0}^{r} \left[ \frac{1}{2} \int_{0}^{R_{i}} \chi \l(r'',r'\r)\l(\frac{\mathrm{d}M}{\mathrm{d}r'}\r)\mathrm{d}r' \right]r''^{2}\mathrm{d}r'' - A \frac{G M}{r^2}+ H_{0}^{2}\Omega_{\Lambda}^{0} r. \label{Force}
\end{equation}

\noindent Setting $\alpha=0$ and $A=1$ causes the first term of the right hand side (r.h.s) to vanish and the usual force is recovered. This modified force not only implements a time dependence transition between standard gravity at larger scales and A-times stronger gravity at smaller scales, but also introduces an important phenomenological effect that comes from the convolution operator in the Poisson equation.  This operator triggers off the mixing of scales at some physical radius i.e., the force at some physical position $r$ not only depends on the matter inside that radius but also on the configuration of the matter outside.  Although, this scale dependence is weak because of the $\mu$ function is quite cusped at the origin, this breaks the Birkhoff's theorem and introduces a series of effects that alter the way an spherical pulse collapses, and therefore, a different value for the density threshold $\delta_{c}$ is now obtained. The main reason behind this is that the collapse now depends on the configuration (size and shape) of the density profile due to the scale dependent new force. Given this, an initial homogeneous matter density evolves to a non-constant profile that eventually may suffer from shell-crossing. This the reason why we assumed a decreasing density profile (\ref{gauss1}), for which we have to follow the evolution of each concentric shell in order to avoid shell-crossing before the turn-around time. A similar situation is studied in \cite{Martino:2008ae}, where a Yukawa force is considered. \\

Now we proceed to solve numerically  (\ref{newton}) to find the radius of each inner shell as function of the scale factor, $r \l(a\r)$.  To accomplish this, it is convenient to make a change variable $r \equiv \frac{r_i}{a_i}x$, as in \cite{Rubin:2013sfa}, in which $r_{i}$ is the initial position of the corresponding shell and $a_{i}$ is the value on the scale factor at initial time, that we set to be $a_{i}=0.01$ ($z\approx 100$) that is well inside matter-dominant era and plenty before the accelerated cosmic expansion ($z < 1$). Also, we set the perturbation amplitude $\delta_i \approx 10^{-2}$, to have initial conditions within the linear theory. Taking into account the results of the previous section, the equation of motion becomes
\begin{equation}
\frac{\mathrm{d}^ {2}x}{\mathrm{d}t^{2}}= - \frac{4 \pi G}{x^2} \frac{a_{i}^{3}}{r_{i}^{3}} \int_{0}^{x} \left[ \frac{1}{2} \int_{0}^{X_{i}} \chi \l(x'',x'\r)\l(\frac{\mathrm{d}M}{\mathrm{d}x'}\r)\mathrm{d}x' \right]x''^{2}\mathrm{d}x'' - \frac{1}{2} A \frac{H_{0}^{2} \Omega_{m}^{0}}{x^{2}}\l(1+\bar{\delta}_{i}\r) + H_{0}^{2}\Omega_{\Lambda}^{0} x  . \label{ForceAdi1}
\end{equation}

\begin{figure}
\begin{center}
\includegraphics[scale=.6]{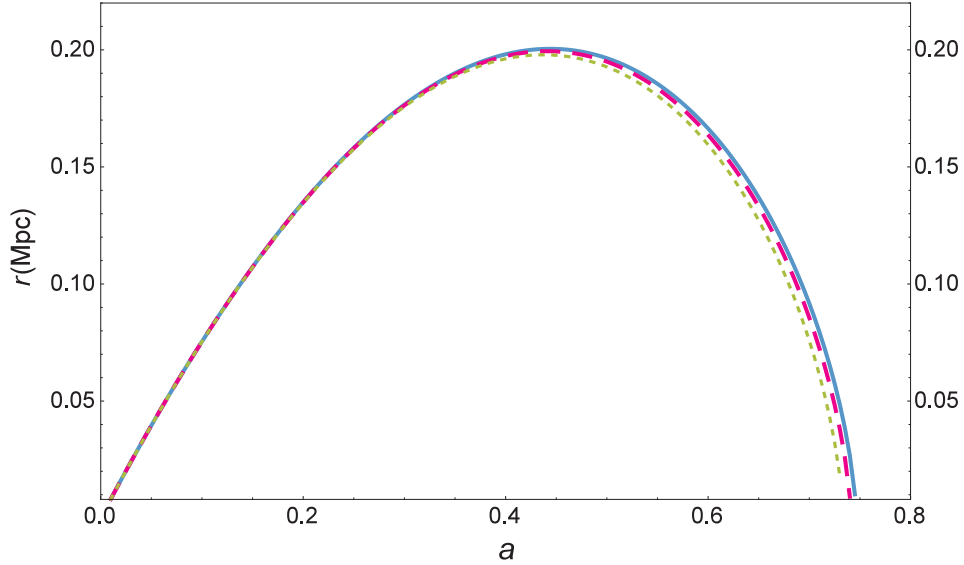}  
\caption{Evolution of the innermost shell in different gravity models. In blue (solid) the $\Lambda$CDM model with standard gravity; in magenta (dashed) and in green (dotted) MG gravity models with $\alpha=-0.001$ and $\alpha=-0.0025$, respectively, keeping $\beta_{0}=1\hspace{.1cm} \mathrm{Gpc}^{-1}$ in both models.} \label{imagen}
\end{center}
\end{figure}

\noindent Integrating by parts the first term of the r.h.s one gets:

\begin{eqnarray}\nonumber
\frac{\mathrm{d}^ {2}x}{\mathrm{d}t^{2}}&=& - \frac{2 \pi G}{x^2} \frac{a_{i}^{3}}{r_{i}^{3}} \l[M\l(X_{i}\r) \int_{0}^{x} \chi \l(x'',X_{i}\r) x''^{2}\mathrm{d}x'' - \int_{0}^{X_{i}} M\l(x'\r) \int_{0}^{x} \l(\frac{\mathrm{d}\chi \l(x'',x'\r)}{\mathrm{d}x'}\r) x''^{2}\mathrm{d}x'' \mathrm{d}x'\r]\\  &&- \frac{1}{2} A \frac{H_{0}^{2} \Omega_{m}^{0}}{x^{2}}\l(1+\bar{\delta}_{i}\r) + H_{0}^{2}\Omega_{\Lambda}^{0} x  , \label{ForceAdi}
\end{eqnarray}

\noindent where we defined $X_{i} \equiv \frac{a_i}{r_i}R_i$.\\

To perform the integration we have to specify the MG model, given by (\ref{nppf2}). We used the BZ model, eq. (\ref{mubz}), to determine (\ref{mu}), in which	
\begin{eqnarray}
\alpha & = & 1-A, \\ 
\beta^2 & = & \beta_{0}^{2}a^{-4}=\lambda_{1}^{-2}a^{-4},\label{refe}
\end{eqnarray}

\noindent in which we have taken the values of $A$ that imply small deviations from GR and we set $\lambda_{1}=1 \hspace{.1cm}\mbox{Gpc}$. We have assumed a background $\Lambda$CDM model with no curvature and $\Omega_{m}=0.31$. 

\noindent In the figure (\ref{imagen}) we show $x\l(a\r)$ for the innermost shell for $\Lambda$CDM model ($\alpha=0$) and for two particular MG models with $\alpha=-0.001$ and $\alpha=-0.0025$.  As it can be seen from the figure each shell with initial radius $r_i$ expands until it reaches a maximum and then turns around to eventually collapse as $r\rightarrow 0$. For $\alpha < 0$ the gravity is stronger than the $\Lambda$CDM model causing the shells to collapse faster and for $\alpha>0$ a slower collapse takes place. One refers as the turn around time ($t_{\mathrm{TA}}$) to the time at which the outermost shell reach its maximum expansion. After turn around the collapse follows but, physically, the perturbation will never reach the singularity due to merging or violent relaxation that establish a virial equilibrium. Therefore, it is customary to assume that the density contrast of the dark matter halo virializes at twice the turn-around time. From figure (\ref{imagen}) it is clear the known effect of a fifth force to increase the gravitational potential to develop an earlier collapse than the $\Lambda$CDM model. \\

\begin{figure}
\begin{center}
\includegraphics[scale=.6]{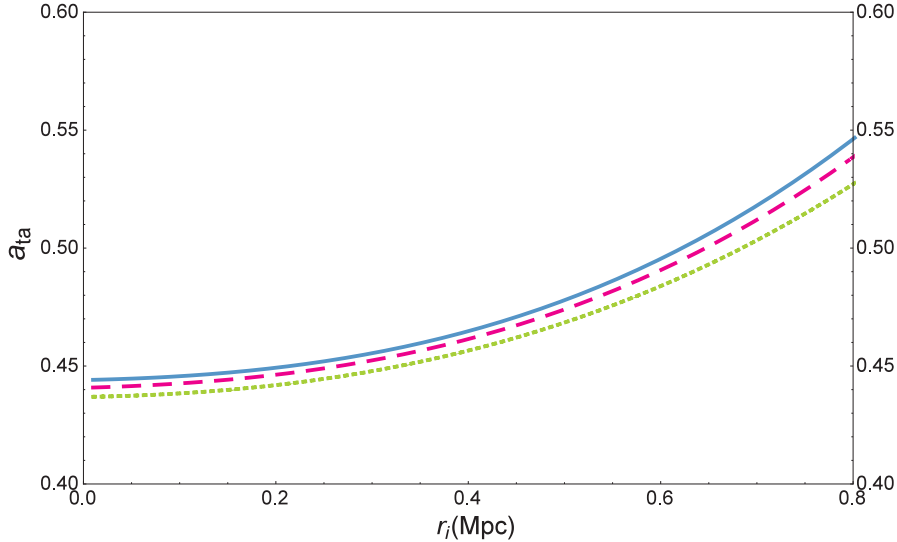}  
\caption{Scale factor at turn around as function of the initial shell position, showing that there is no shell crossing, at least not before the turn around time. In blue (solid) $\Lambda$CDM model; in magenta (dashed) and in green (dotted) MG gravity models with $\alpha=-0.001$ and $\alpha=-0.0025$, respectively, keeping $\beta_{0}=1 \hspace{.1cm} \mathrm{Gpc}^{-1}$ in both models.} \label{img2}
\end{center}
\end{figure}

Because of a non-constant initial density profile the evolution equation for the inner shells are no longer self-similar. For instance, in the top-hat case each shell evolution equation is indeed self-similar, meaning that each inner shell reaches its turn-around point (and its collapse point) at the same time.  In the case of a monotonically decreasing initial density profile, as figure (\ref{img2})  shows,  the time at which each inner shell reaches its turn-around point (and its collapse point) is a monotonically increasing function of the initial shell position $r_{i}$. For the sake of not spoiling the SC model we need to set the edge of the initial Gaussian profile so that the outermost shell reaches the turn-around point before the innermost collapses. In general, this can be achieved demanding $\bar{\delta}_{i}\l( 0\r)/ \bar{\delta}_{i}\l(R_{i}\r)\leq 1.587$ \cite{Rubin:2013sfa}, where $R_i$ is the radius of the initial density contrast. In the case of a Gaussian profile this condition  can be easily accomplished by setting the edge of the Gaussian pulse at $R_i =0.8 \lambda^{-1}$, see eq. (\ref{gauss1}).\\

\begin{figure}
\begin{center}
\includegraphics[scale=.8]{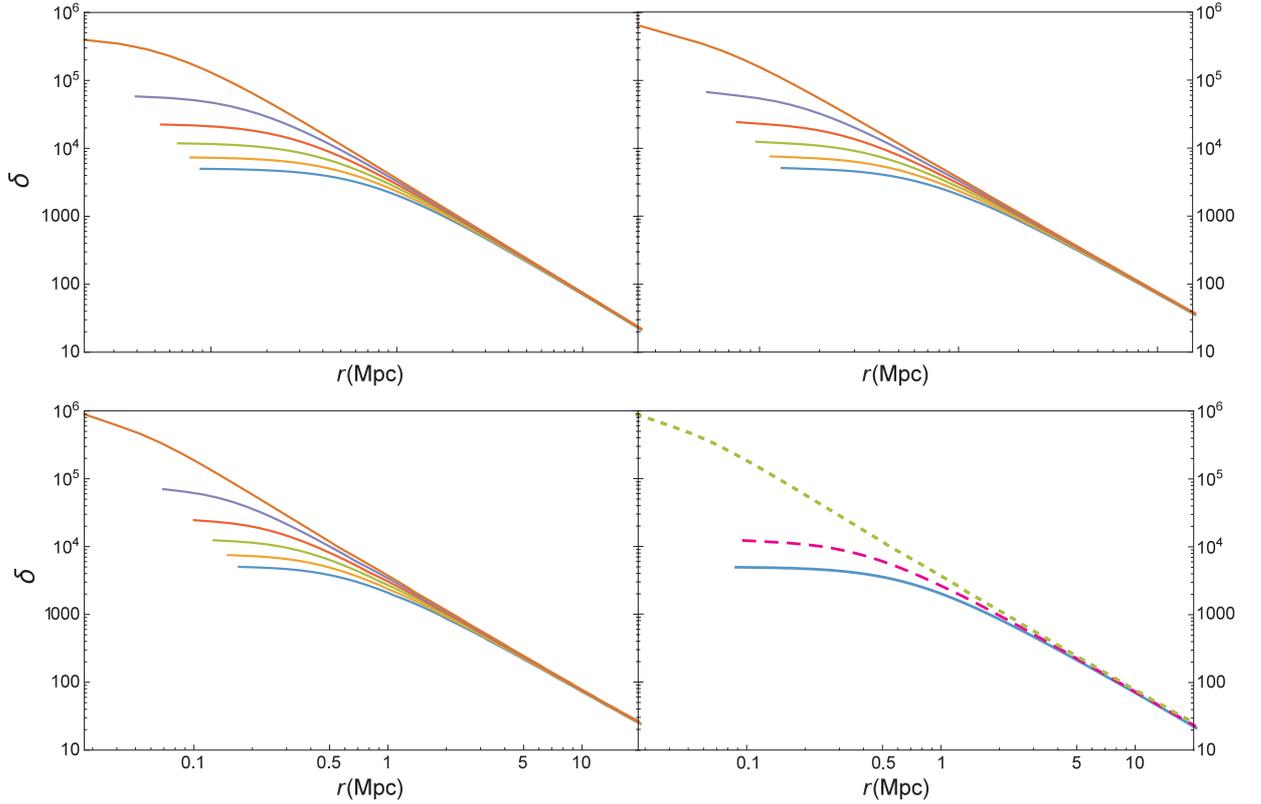}
\caption{The latest time steps of the density contrast $\delta$ evolution before it reaches the collapse criteria for the innermost shell. Top-left: $\Lambda$CDM model; top-right: MG model with $\alpha=-0.001$ and $\beta_{0}=1 \hspace{.1cm} \mathrm{Gpc}^{-1}$; bottom-left: MG model with $\alpha=-0.0025$ and $\beta_{0}=1 \hspace{.1cm}\mathrm{Gpc}^{-1}$; bottom-right: The previous three models at the same time (the time at which the model with $\alpha=-0.0025$ collapses). In blue (solid) the $\Lambda$CDM model, in magenta (dashed) and in green (dotted) the MG models with $\alpha=-0.001$ and $\alpha=-0.0025$, respectively.}\label{colap2}
\end{center}
\end{figure}

Figure (\ref{colap2}) shows the evolution of density contrasts a few time steps before the innermost shell collapses; it is the latest time the SC model is still valid. The top-left panel shows the $\Lambda$CDM model, the top-right and bottom-left panels show MG gravity models with $\alpha=-0.001$ and $\alpha=-0.0025$, respectively, keeping $\beta_{0}=1 \hspace{.1cm} \mathrm{Gpc}^{-1}$ in both models, see eq. \ref{refe}. As it can be notice from the figure the three models ($\alpha=0$,$-0.001$,$-0.0025$) have a similar behavior with the only difference in the collapsed times. In the MG cases the collapse is slightly faster than in $\Lambda$CDM model (or slower if $\alpha>0$). This can be seen in the bottom-right panel where the three models are plotted at the same time (the time at which the innermost shell of the model with $\alpha=-0.0025$ collapses). As expected, the gravity model with a stronger deviation from GR collapses faster, thus, $\delta$ is bigger. Despite the differences among models, when they are plotted at their collapsed time (the time at which their innermost shell collapses) the final behaviors look more similar, finding the same pattern: a model with stronger deviation has a denser profile, according to the $\alpha$ parameter. But the small differences shown in figure (\ref{colaptime}) cause considerable changes in the density threshold $\delta_{c}$, as it is shown below.\\

\begin{figure}
\begin{center}
\includegraphics[scale=.6]{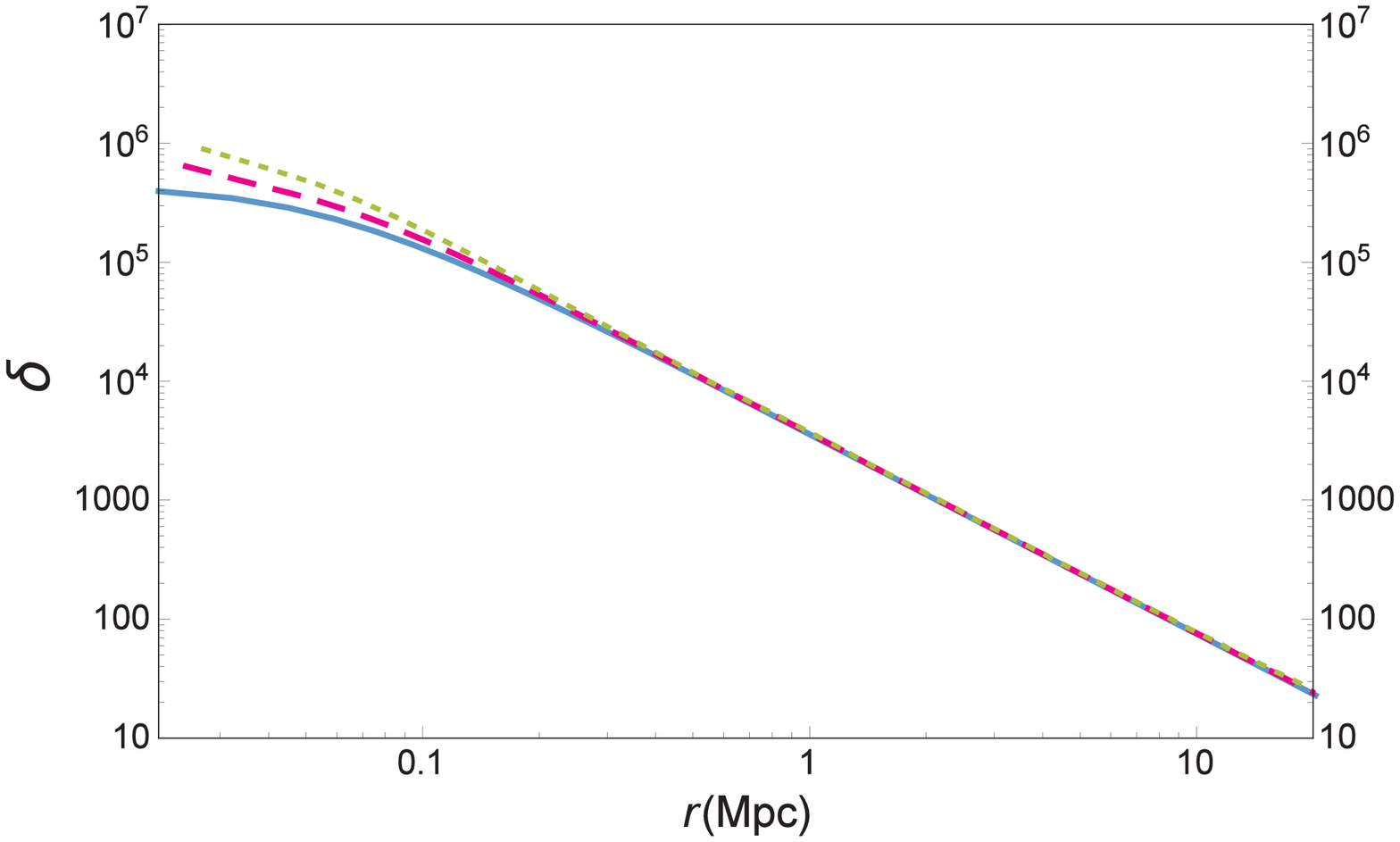}
\caption{The density contrasts $\delta$ at their collapse time, given the corresponding criteria for the innermost shell.  In blue (solid) the $\Lambda$CDM model, in magenta (dashed) and in green (dotted) MG models with $\alpha=-0.001$ and $\alpha=-0.0025$, respectively.}\label{colaptime}
\end{center}
\end{figure}

\begin{figure}
\begin{center}
\includegraphics[scale=.8]{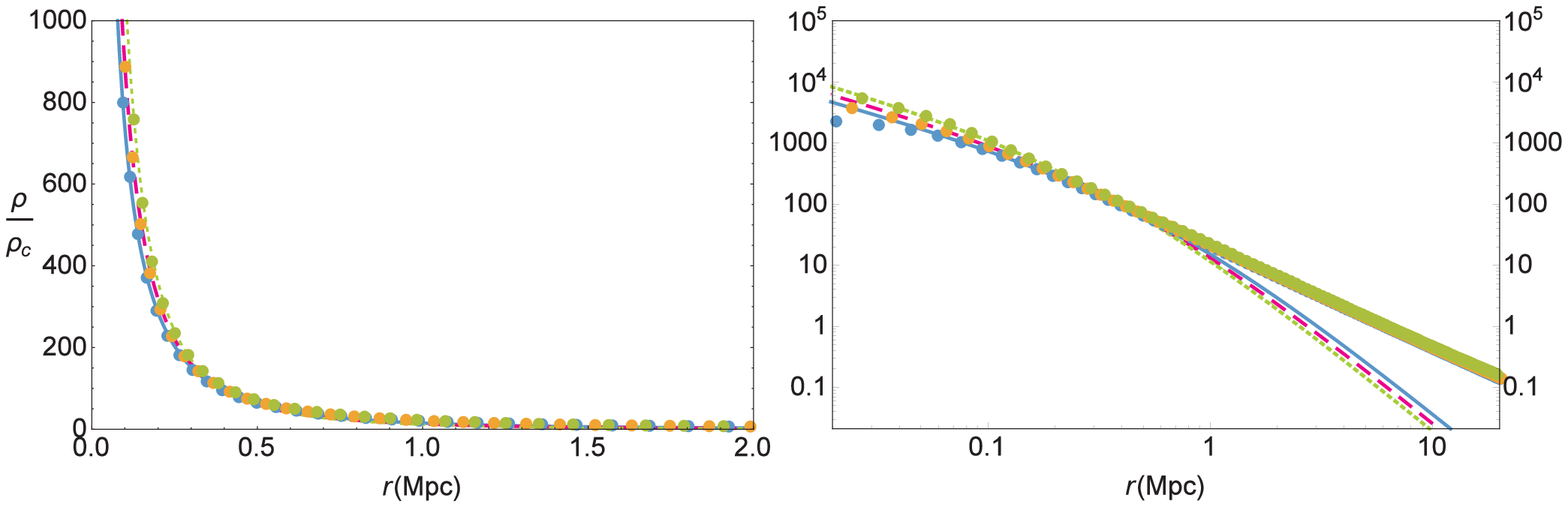}
\caption{The density profiles (over the critical density today). Left panel: in blue the $\Lambda$CDM model, in magenta and in green MG models with $\alpha=-0.001$ and $\alpha=-0.0025$, respectively.  Dots represent density profiles obtained from the SC model at their collapse criteria for the innermost shell and in solid lines NFW fitting profiles with $r_{v}= 3.19$ and $c = 4.95$; $r_{v}= 2.99$ and $c = 6.52$; $r_{v}= 2.86$ and $c = 8.60$, respectively. Right panel: the same as in left panel but in log scale.}\label{fnfw}
\end{center}
\end{figure}

The final profiles we have obtained do not represent virialized structures, because their outermost shells have just past the turn around time and they have not reached a virial condition. Following, we test how much the collapsed structures differ from a typical collapse profile. Accordingly, we fit a Navarro-Frenk-White (NFW) profile \cite{Navarro:1995iw} to our final profile, which is given by:

\begin{equation}
\rho = \frac{\rho_{0}}{c \frac{r}{r_{v}}\l(1+c \frac{r}{r_{v}}\r)^2},\label{nfw}
\end{equation}

\noindent where $c$ is called the concentration parameter and $r_{v}$ is the radius of the virialized halo, and their values vary from halo to halo; $\rho_{0}$ is a typical halo density that depends on $r_{v}$, $c$, and cosmological parameters. Figure (\ref{fnfw}) shows the numerical density profiles obtained (dots), divided by the critical density evaluated at $a=1$; and the NFW fittings (solid). As it can be seen from the figure, right panel, the fittings are very close to the density profiles (obtained from the SC model) in the innermost shells, but they start to deviate from $r\approx 1 $Mpc. This is because the outermost shells of each pulse are in their way to collapse and have not reached a virial condition yet.\\

\noindent Furthermore the collapse time, or the time at which the halo virializes, depends on the initial value of the density contrast $\delta_{0}$; the higher $\delta_{0}$ the earlier the overdense pulse 
collapses. The value of the linear density contrast extrapolated at the time of collapse is referred as the critical threshold $\delta_{c}$. This quantity is important because it represents a key 
element in calculating the mass function \cite{Press:1973iz}, the number density $\mathrm{d}n$ of collapse halos with mass in the $\mathrm{d}M$ range. The analytic expression is given for $\delta_{c}$ in the case of the EdS model \cite{white:2008}:

\begin{equation}
\delta_{c}\l(a_{c}\r) = \frac{3}{5}\pi ^{2/3}\l[H\l(a_{c}\r)t_{c}\r]^{-2/3},
\end{equation}

\noindent and for the $\Lambda$CDM model

\begin{equation}
\delta_{c}\l(a_{c}\r) = \frac{3}{5} g\l(a_{c}\r) \frac{a_i}{X_\mathrm{ta}}\l(1+\frac{\Omega_{\Lambda}^{0}}{\Omega_{m}^{0}}\frac{X_{\mathrm{ta}}^3}{1+\bar{\delta}_{i}}\r)\l(1+\bar{\delta}_{i}\r)^{1/3}. \label{deltac}
\end{equation}

\begin{figure}
\begin{center}
\includegraphics[scale=.8]{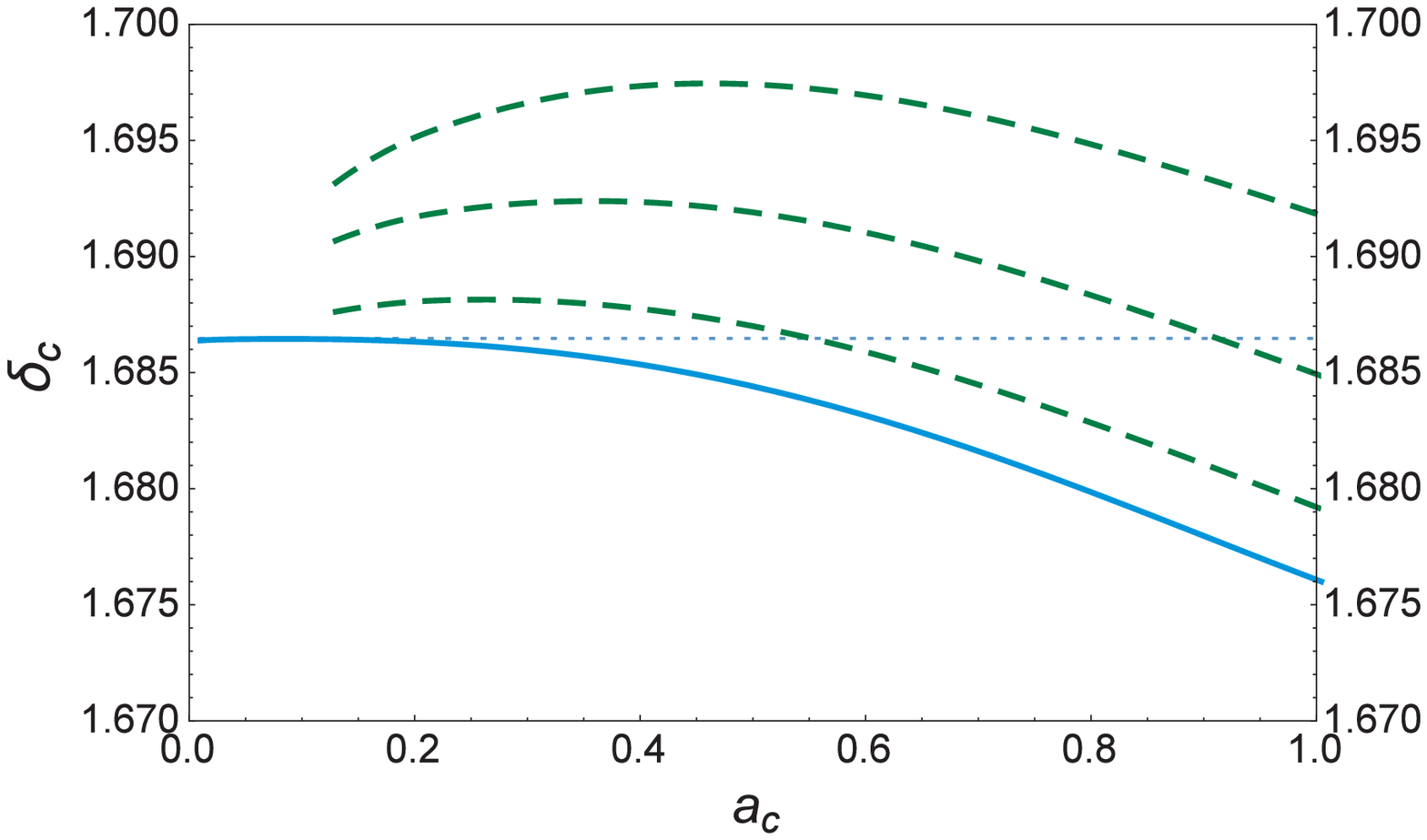}
\caption{Linear density threshold at collapsing time, $\delta_{c}$, for different models. In dotted (blue) the EdS model, in solid (blue) the $\Lambda$CDM model and in dashed lines (green) MG models with different $\alpha$ values, from bottom to the top: -0.001, -0.0025 and -0.005. As $\alpha$ decreases its values, $\delta_{c}$ departs more from the $\Lambda$CDM model. We set $\beta_{0}=1\hspace{.1cm} \mathrm{Gpc}^{-1}$ in all MG models.}\label{imagen3}
\end{center}
\end{figure}

\noindent $g\l(a_{c}\r)$ is the linear growth factor evaluated at collapse time.  In MG models one can use the same expression for $\delta_{c}$ as equation (\ref{deltac}), but in this case $g\l(a\r)$ is the linear growth factor for the modified linear differential equation for the evolution of $\delta$ in the PPF formalism, which in Fourier space is given by

\begin{equation}
\delta '' + \mathcal{H}\delta '-\frac{3}{2}\mathcal{H}^{2}\Omega_{m}\mu\l(a,k\r) \delta =0, \label{linear}
\end{equation}

\noindent where $'$  denotes derivation with respect conformal time. This last equation follows from the energy-momentum tensor conservation, considering MG only enter through the Poisson's equation.\\

\noindent Using the BZ parametrized MG model (\ref{mubz}), $\lambda_1$ is the only free parameter which can be related to the $f\l(R\r)$ dimensionless parameter $B_{0}$ \cite{Song:2006ej}:\\

\begin{equation}
B_{0}\equiv \frac{f_{R}}{f_{RR}}=\frac{2 H_{0}^{2}}{c^{2}}\lambda_{1}^{2} = \frac{2 H_{0}^{2}}{c^{2}} \beta_{0}^{-2}.
\end{equation}

\noindent In \cite{Marchini:2013lpp} the authors obtained an upper limit to this dimensionless length scale $B_{0} < 0.14$ at 95\% c.l. from CMB measurements. From this bound we have that $\beta_{0} > 1 \hspace{.1cm}\mathrm{Gpc}^{-1}$, which means that in scales larger than  $\lambda_1 \approx 1$ Gpc one has to recover GR.\\

In figure (\ref{imagen3}) results are shown for $\delta_{c}\l(a_c\r)$ varying the parameter $\alpha$. The usual constant result (1.686) for an EdS Universe is shown in dotted line (blue) and in solid line (blue) for the $\Lambda$CDM model in accordance with known results \cite{Lokas:2001nw}. For the MG models presented here we note that $\delta_{c}\l(a_c\r)$ tends to the $\Lambda$CDM case as $\alpha \rightarrow 0$. As $\alpha_{0}$ decreases (starting from zero) $\delta_{c}\l(a_c\r)$ deviates from the $\Lambda$CDM case and for much larger values it gets over its EdS counterpart.

\section{Conclusions}

We have worked out the SC in MG models of the PPF type. We found some analytic expressions for a general modification of gravity of type given by equation (\ref{poissonmg}) and found numerical solutions for the parametrization designed to mimic the $f\l(R\r)$ gravity behaviour proposed by Bertschinger and Zukin (BZ). As it happens in $f\l(R\r)$ models, an initial constant profile suffers from shell-crossing, making our analysis invalid. Therefore, in order to avoid shell-crossing a Gaussian density profile was proposed, but in general a decreasing profile works as well; figure (\ref{img2}) shows shell-crossing is avoided in the models presented in this work. We solved with some detail the evolution of two different MG models with $\alpha=-0.001$ and $\alpha=-0.0025$, respectively, keeping the MG transition scale fixed to $\beta_{0}\equiv \lambda_{1}^{-1}=1 \hspace{.1cm}\mathrm{Gpc}^{-1}$. In figure (\ref{colap2}) and in figure (\ref{colaptime}) show, as it was expected, gravity models with strongest deviation from GR collapses most and fastest.\\

Though the final density profiles obtained do not represent virialized structures, we fitted a NFW profile to the each of them. In figure (\ref{fnfw}) we showed that the NFW fittings are very close to the final density profiles (obtained from the SC model) in their innermost shells, but they start to deviate for the most external shells. This is because the outermost part of the pulses have just passed the turn around point and they are in the way to reached a virial condition. It is interesting that NFW profiles perform good fits to MG collapse models, at least at the stage considered in this work.\\

Finally, figure (\ref{imagen3}) shows the density threshold $\delta_{c}$ for different MG models. These can deviate considerably from the $\Lambda$CDM behaviour depending upon the values of the $\alpha$ parameter, which dictates the strength of the fifth force. We kept the value of $\beta_{0}$ fixed to $\beta_{0}=1\hspace{.1cm}\mathrm{Gpc}^{-1}$ which determines the time transition between standard gravity force at early times and A-times stronger gravity at recent times. Further, we noted that the fifth force introduced does not follow the Birkhoff's theorem making the spherical collapse model to depend also on the density profile configuration. Our results indicate that the collapsed objects are denser than $\Lambda$CDM predictions and this effect strengthens at smaller redshifts.\\

\bibliographystyle{unsrt}

\end{document}